\documentclass[sigconf, nonacm]{acmart}

\AtBeginDocument{%
  \providecommand\BibTeX{{%
    \normalfont B\kern-0.5em{\scshape i\kern-0.25em b}\kern-0.8em\TeX}}}

\copyrightyear{2021}
\acmYear{2021}
\setcopyright{acmlicensed}\acmConference[SIGIR '21]{Proceedings of the 44th International ACM SIGIR Conference on Research and Development in Information Retrieval}{July 11--15, 2021}{Virtual Event, Canada} \acmBooktitle{Proceedings of the 44th International ACM SIGIR Conference on Research and Development in Information Retrieval (SIGIR '21), July 11--15, 2021, Virtual Event, Canada}
\acmPrice{15.00}
\acmDOI{10.1145/3404835.3462842} \acmISBN{978-1-4503-8037-9/21/07}

\usepackage{booktabs}
\usepackage{multirow}
\usepackage{marvosym}
\usepackage{titlesec}
\usepackage{balance}

\titlespacing*{\subsection}
  {0pt}{0.5\baselineskip}{0pt}
\titlespacing*{\section}
  {0pt}{0.5\baselineskip}{0pt}
\titlespacing*{\subsubsection}{0pt}{0.5\baselineskip}{2pt}
\begin{document}
\fancyhead{}
\setlength{\abovedisplayskip}{0pt}
\setlength{\belowdisplayskip}{0pt}
\title{A General Method For Automatic Discovery of Powerful Interactions In Click-Through Rate Prediction}

\author{Ze Meng$^{1,\star,\dagger}$, Jinnian Zhang$^{2,\star,\dagger}$, Yumeng Li$^{3}$, Jiancheng Li$^{3}$, Tanchao Zhu$^{3}$, Lifeng Sun$^{{4,5},{\dagger}}$}
\authornote{Both authors contributed equally to this research.}
\authornote{Corresponding authors.}
\affiliation{$^{1}$Beijing Key Lab of Networked Multimedia, Dept. of Computer Science and Technology, Tsinghua University \country{China}}
\affiliation{$^{2}$Dept. of Electrical and Computer Engineering, University of Wisconsin Madison \country{USA}}
\affiliation{$^{3}$Alibaba Group \country{China}}
\affiliation{$^{4}$Key Laboratory of Pervasive Computing, Tsinghua University, Ministry of Education \country{China}}
\affiliation{$^{5}$BNRist, Dept. of Computer Science and Technology, Tsinghua University \country{China}}
\affiliation{$^{\dagger}$mengz18@tsinghua.org.cn, jinnian.zhang@wisc.edu \country{sunlf@tsinghua.edu.cn}}

\renewcommand{\shortauthors}{Meng et al.}
\renewcommand{\authors}{Ze Meng, Jinnian Zhang, Yumeng Li, Jiancheng Li, Tanchao Zhu, Lifeng Sun}

\begin{abstract}
  Modeling powerful interactions is a critical challenge in Click-through rate (CTR) prediction, which is one of the most typical machine learning tasks in personalized advertising and recommender systems.
Although developing hand-crafted interactions is effective for a small number of datasets, it generally requires laborious and tedious architecture engineering for extensive scenarios.
In recent years, several neural architecture search (NAS) methods have been proposed for designing interactions automatically.
However, existing methods only explore limited types and connections of operators for interaction generation, leading to low generalization ability.
To address these problems, we propose a more general automated method for building powerful interactions named AutoPI.
The main contributions of this paper are as follows: AutoPI adopts a more general search space in which the computational graph is generalized from existing network connections, and the interactive operators in the edges of the graph are extracted from representative hand-crafted works. It allows searching for various powerful feature interactions to produce higher AUC and lower Logloss in a wide variety of applications.
Besides, AutoPI utilizes a gradient-based search strategy for exploration with a significantly low computational cost. Experimentally, we evaluate AutoPI on a diverse suite of benchmark datasets, demonstrating the generalizability and efficiency of AutoPI over hand-crafted architectures and state-of-the-art NAS algorithms.
\end{abstract}



\maketitle

\section{Introduction}
Predicting the user's clicking probability on a given item is a crucial task in a recommender system. 
Essentially, the key challenge in CTR prediction is how to model powerful feature interactions effectively.
The classical methods employ a single type of interactive operator to model the fixed-order feature interactions in an explicit manner. For example, Factorization Machines (FM) based methods \cite{rendle2012factorization, xiao2017attentional} learn 2nd-order cross features using inner product. And its variant, High-order Factorization Machine (HOFM) \cite{blondel2016higher}, adopted Analysis of Variance (ANOVA) to approximate high-order cross features, while only marginal improvement was observed despite using more parameters.
Deep learning introduces a more effective operator, Single-layer Perceptron (SLP) \cite{hornik1989multilayer}, which can be stacked to form Multi-layer Perceptron (MLP) to generate high-order feature interactions in an implicit manner.
However, the representation ability of the cross features generated from a single type of operator is limited.
As an enhancement to MLP, incorporating explicit lower-order feature interactions in deep learning models achieves better performance, namely, multi-interaction ensemble, such as Wide\&Deep \cite{cheng2016wide}, DeepFM\cite{guo2017deepfm} and xDeepFM \cite{lian2018xdeepfm}.

However, manually developing the interactions for generalizing across diverse scenarios is computationally intensive and is a black box optimization without an optimality guarantee.
Instead of fixing the order of interactions, learning the order of interactions or adaptively identifying important groups of features within a predefined order attract increasing attention. Specifically, by using the multi-head self-attention \cite{song2019autoint}, the relevant features can be grouped to form meaningful arbitrary-order cross features. AFN \cite{cheng2020adaptive} learns adaptive-order cross features via a logarithmic transformation layer.
And more advanced NAS methods, AutoFIS \cite{liu2020autofis} automatically identify important cross features from all possible 2nd-order and 3rd-order combinations of features with inner product. As a generalization of AutoFIS, AutoGroup \cite{liu2020autogroup} can generate substantial interactions of any order in an automatic manner. AutoCross \cite{luo2019autocross} explicitly search for useful high-order cross features in a tree-based search space by an efficient beam search method. 
However, the above methods generate the interactions by using only a single type of operator for all identified groups of features. In order to enable the automatic search for the best connections of operators for generating the interaction from each group,
AutoFeature \cite{khawar2020autofeature} encodes the subnetwork for each candidate group of features within a predefined order as a string. Note that each subnetwork can be considered as a direct acyclic graph (DAG) of operators. Then a tree of Naive Bayes classifiers is used for optimization.
AutoCTR \cite{song2020towards} defines three basic blocks from operators SLP, FM, and dot product (DP), and utilize the evolutionary algorithm to obtain the best connections of these blocks. 
Similarly, in AutoRec \cite{wang2020autorec}, five different blocks are created based on several operators such as SLP, FM, outer product, self-attention, etc. Three strategies (random, greedy, and Bayesian) are used to optimize the number and type of blocks for ensemble, and the hyperparameters in these blocks.

Nevertheless, these works only involve simple operators and limited possible connection types for building the interactions, leading to narrow search space.
For generalization to a large number of scenes, we study a more general paradigm of discovering powerful interactions.
We search the interactions of CTR prediction according to three key factors: (1) the order of interaction in each tower, (2) the types and connections of operators used to construct interactions, (3) the multi-tower structure for the ensemble of interactions. 
Specifically, we review various representative literature to extract and modulize significant operators to form the operation space and design the tailored cell-based computational graph containing more possible connections of operators. Each node in the graph represents a latent representation (e.g., intermediate cross features), and the edge is an operator from the operation space.
Notably, we unify the concept of the order related to explicit and implicit interactions following \cite{lian2018xdeepfm}.
For example, DeepFM contains two types of interactive operators, FM and SLP.
FM explicitly produces 2nd-order cross features, while MLP implicitly generates 4th-order cross features because it has three SLPs.
Thus, our method discovers powerful interactions by integrating multiple interactions, where each interaction consists of different numbers, types, and connections of operators.
However, our search space exponentially increases with the aforementioned three key factors of the interactions, requiring an efficient search strategy.
Inspired by DARTS \cite{liu2018darts}, we describe an improved gradient-based method for exploration to improve efficiency.
In detail, we introduce architecture parameters to relax the discrete set of operators and node connections, and utilize the Bi-level optimization algorithm to optimize the architecture and weights by gradient descent.
Moreover, two different training techniques are proposed to alleviate a well-known problem in the weight-sharing mechanism, w.r.t. the performance gap \cite{xie2020weight}.

To summarize, the main contributions of this paper can be highlighted as follows:
\begin{itemize}
\item Designing a general and efficient search space to generate powerful interactions, which consists of the operation space and the computational graph. The operation space is constructed via extracting and modulizing the representative interactive operators in existing literature, and we design a tailored cell-based computational graph for finding the best connections of the operators.
\item A gradient-based search strategy is applied for exploration in our broad search space to improve efficiency. We introduce the architecture parameters for relaxing the discrete search space and use a Bi-level optimization algorithm to optimize the architecture and weights iteratively.
\item Two training techniques are utilized for alleviating the performance gap between the continuous architecture encoding and the derived architecture.
\item Extensive experiments demonstrate the generalization and efficiency of AutoPI on datasets in different scales.
Comparing with SOTA methods, on average, our method increases the AUC by $0.46\%$ and $0.70\%$ on the public and private datasets, respectively. Furthermore, the gradient-based search strategy can reduce $90\%$ of the search overhead compared to the NAS baselines.
\end{itemize}
Our work is organized as follows. In Section \ref{sec:rw}, we summarize the related work on NAS and CTR prediction. Section \ref{sec:def} formally defines our problem. 
Section \ref{sec:autopi} presents the details of our method.
In Section \ref{sec:exp}, we present the experimental results and detailed analysis. Finally, we conclude this paper and point out the future work in Section \ref{sec:disc}.

\section{Related Work}
\label{sec:rw}
\textbf{Click-Through Rate Prediction.} Traditional models aim to formulate the explicit fixed-order interactions for CTR prediction, such as LR, FM, Attention Factorization Machine (AFM), HOFM, and CrossNet \cite{qu2018product}.
Different from HOFM, the number of parameters in CrossNet only increases linearly with the input dimension. 

Compared to polynomial interactions in FM, CrossNet, etc., 
complicated non-linear features created through neural networks \cite{qu2018product, he2017neural, cheng2020adaptive, song2019autoint} can be more informative and thus improve the prediction performance. 
PNN \cite{qu2018product} builds the inter-field interactions by using the inner product or outer product and feeds it to an MLP.
NFM \cite{he2017neural} has a similar architecture as PNN, but uses a Bi-Interaction layer to model the inter-field interactions.
Instead of interactions generated by a single model, multi-interaction ensembles are known as the boost of performance by exploiting the advantages of different models.
In Wide\&Deep \cite{cheng2016wide}, it combines LR and Deep Neural Network (DNN), which outperforms each individual model. 
DeepFM \cite{guo2017deepfm}  comprises DNN and FM,
and other ensemble models include Deep\&Cross \cite{wang2017deep} and xDeepFM \cite{lian2018xdeepfm}. 
These models have been verified to be effective on many public datasets.
In recent years, some hand-crafted models try to learn the order of feature interactions. For example, AutoInt \cite{song2019autoint} makes use of the latest techniques, attention and residual networks, to generate non-linear features. AFN \cite{cheng2020adaptive} learns the powers (orders) of each feature in interactions by a logarithmic transformation layer.

\textbf{Neural Architecture Search.}
The primary motivation for NAS is to automate the laborious process of designing neural networks in different tasks. 
There are basically four existing frameworks for NAS \cite{elsken2019neural}: evolution-algorithm based NAS \cite{real2019regularized}, reinforcement-learning based NAS \cite{pham2018efficient}, Bayesian optimization based NAS \cite{elsken2019neural}, and gradient-based NAS \cite{ren2020comprehensive}.
Among various NAS methods, DARTS \cite{liu2018darts} and its variants \cite{xu2019pc, liang2019darts+, chu2020fair, chu2020noisy, li2020pd}
have become the most popular algorithms due to remarkable efficiency improvement in time cost.
The core of DARTS is to utilize continuous relaxation of the architecture representation and apply 
Bi-level optimization to iteratively update the architecture and its weights by using gradient descent.
The effectiveness of DARTS has been proved in both Computer Vision (CV) and Neural Language Processing (NLP) domains.
Besides existing literature on NAS methods are focused on the tasks in CV and NLP, there is an increasing interest in applying NAS to recommender systems in recent years, such as AutoCTR, AutoFeature and AutoRec, etc.
These methods are demonstrated to yield higher performance than human-crafted models.

\section{Problem Definition}
\label{sec:def}
We summarize the general framework of CTR prediction in Figure \ref{fig:overview}. The deep CTR prediction models have three fundamental stages from a bottom-up perspective: (i) input transformation, (ii) modeling diverse feature interactions, and (iii) multi-interaction ensemble. We formally introduce the definitions of three subproblems in CTR prediction as follows:
\begin{figure}[]
  \centering
  \includegraphics[width=\linewidth]{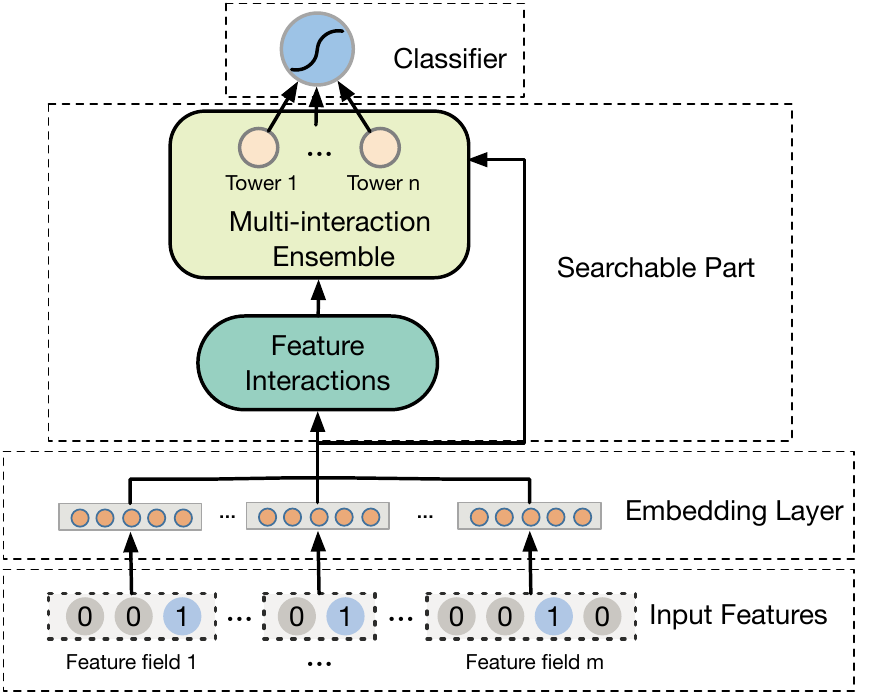}
  \caption{The CTR model consists of the fixed part (embedding layer and classifier) and the searchable part (feature interactions and multi-interaction ensemble).}
  \vspace{-15pt}
  \label{fig:overview}
\end{figure}

\text{DEFINITION} 1. \textbf{(Input Transformation)} 
We represent the raw feature vector as $\textbf{x}=[\textbf{x}_1;\textbf{x}_2;\dots;\textbf{x}_m]$, where $m$ is the number of total fields, $\textbf{x}_{i}$ is a one-hot feature representation of the $i$-th field, 
which is categorical (e.g., gender=male, name=Alan, age=25). Since the feature representations are very sparse and high-dimensional, we employ embedding layers to convert the sparse feature into a low dimensional and dense real-value vector $\textbf{e}_i=\textbf{V}_i\textbf{x}_i$, where $\textbf{V}_i$ is an embedding matrix for field $i$.
Then, the output of the embedding layer will be a feature matrix $\textbf{E}\in\mathbb{R}^{m\times k}$ by stacking multiple embedding vectors $[\textbf{e}_1,\textbf{e}_2,\dots,\textbf{e}_m]$, where $k$ is the embedding size of each field. 
To avoid interference between the deep and shallow model \cite{cheng2020adaptive}, we adopt dual-embedding\footnote{For ease of reading, we omit the dual-embedding in Figure 1.} (i.e., two independent embedding layers $\textbf{V}^{\text{low}}$ and $\textbf{V}^{\text{high}}$) to build lower-order and higher-order feature interactions, respectively. 

\text{DEFINITION} 2. \textbf{(Feature Interactions)} 
The key problem is to determine which types and orders of interactions should be built to form meaningful cross features. We first redefine the order of interactions from the convention following \cite{lian2018xdeepfm}.
Mathematically, Given an input feature matrix $\textbf{E}$, a \textit{p-order cross feature} is defined as: 
\begin{align}
    \textbf{E}^p=o^{(p-1)}(o^{(p-2)}(\dots(o^{(1)}(\textbf{E}))\dots)). 
\end{align}
Namely, the p-order cross feature $\textbf{E}^{p}$ is generated by the composition of \textit{p-1} operators $o^{(1)}(\cdot), o^{(2)}(\cdot), \dots, o^{(p-1)}(\cdot)$ in order, such as FM and SLP. 
Traditionally, this is designed by domain experts based on their knowledge. In this paper, we tackle this problem with a novel method, i.e. NAS.

\text{DEFINITION} 3. \textbf{(Multi-interaction Ensemble)} In hand-crafted CTR prediction models, one of the most critical designs is the multi-interaction ensemble, which combines various orders and types of interactions to form a multi-tower structure. 
For final prediction, the cross features of all towers are concatenated and then delivered to an SLP as follows: 
\begin{align}
\hat{y}=\sigma\left(\textbf{w}^{\intercal}\left(\textbf{e}^{(1)} \oplus \textbf{e}^{(2)} \oplus \cdots \oplus \textbf{e}^{(n)}\right)+b\right),
\end{align}
where $\textbf{w}$ is a column projection vector which linearly combines concatenated features, and $b$ is the bias. Symbol $\oplus$ is the concatenation operation, and  $\sigma(x)=1/(1+e^{-x})$ transforms the logits to users' clicking probabilities. Note that $\textbf{e}^{(i)}$ is the flattened cross features of $\textbf{E}^{(i)}$ from the $i\text{-th}$ tower, and $n$ is a hyperparameter that indicates the total number of towers. The $\textit{Logloss}$ is adopted for training the model, which is defined as follows:
\begin{align}
\textit{Logloss}=-\frac{1}{N} \sum_{i=1}^{N}\left(y_{i} \log \left(\hat{y}_{i}\right)+\left(1-y_{i}\right) \log \left(1-\hat{y}_{i}\right)\right),
\end{align}
where $y_i$ and $\hat{y}_i$ are the ground truth of user clicks and predicted CTR, respectively, and
$N$ is the number of training samples. 
\section{AutoPI}
\label{sec:autopi}
\subsection{Method Overview}
\label{sec:bigpic}
\begin{figure}[h]
  \centering
  \includegraphics[width=\linewidth]{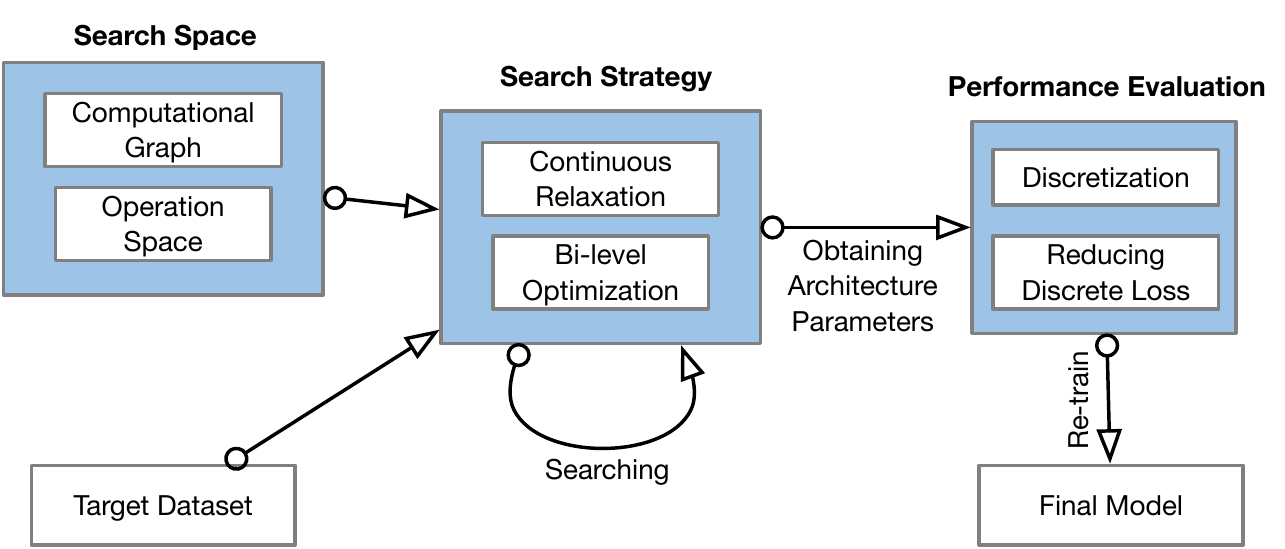}
  \caption{Method overview of AutoPI.}
  \label{fig:method}
\end{figure}

The work flow of AutoPI is shown in Figure \ref{fig:method}. Roughly speaking, our method includes three parts: (i) search space (Section \ref{sec:ssd}), (ii) search strategy (Section \ref{sec:st}) and (iii) performance evaluation (Section \ref{sec:pe}).
Specifically, we adopt the cell-based search space, which consists of the computational graph (Section \ref{sec:graph}) and the operation space (Section \ref{sec:op}).
The efficient gradient-based search strategy is utilized to search for the best connections of operators. Its core is a continuous relaxation (Section \ref{sec:cr}) scheme for our discrete search space, leading to a differentiable learning objective for the joint optimization of the architecture and its weights (Section \ref{sec:opt}).
We also apply the Batch Normalization \cite{ioffe2015batch} technique, because of its capability of dealing with numerical instability and improving fairness among operators (Section \ref{sec:bn}).
After the search process, the discretization technique (Section \ref{sec:dis}) is used to obtain the discrete architecture from continuous architecture encoding. 
For alleviating the performance gap before and after discretization, 
temperature anneal and noisy skip-connection can be utilized in the search process (Section \ref{sec:rdl}).
The final procedure is training the model weights with the powerful architecture from scratch.
\subsection{Search Space Design}
\label{sec:ssd}
We summarize the interactions of hand-crafted models as three critical factors:
(1) The orders of interactions, w.r.t, how many operators should be combined to form an interaction.
(2) Operators used to construct interactions, w.r.t, which operators should be selected and how operators are connected to build an interaction.
(3) Multi-interaction ensemble, w.r.t, which interactions should be integrated for final prediction.
Therefore, our search space consists of the operation space and tailored cell-based computational graph. Each node in the graph represents for a latent representation (i.e., an arbitrary-order cross feature), and the edge is an operator from the operation space.
For better generalization to various applications, the search space should contain sufficient nodes in the computational graph, while potentially encompassing distinct interactive operators in the operation space.
And the multi-tower structure is contained in the graph for the multi-interaction ensemble to improve the performance. Note that the connections of operators in the multi-tower structure are also automatically optimized. 
As a result, we extract diverse interactive operators from representative hand-crafted works to form the operation space. Moreover, two types of computational cells are designed as building blocks of the graph, including an interaction cell and an ensemble cell. 
\subsubsection{Computational Graph.}
\label{sec:graph}
\begin{figure}[]
  \centering
  \includegraphics[width=\linewidth]{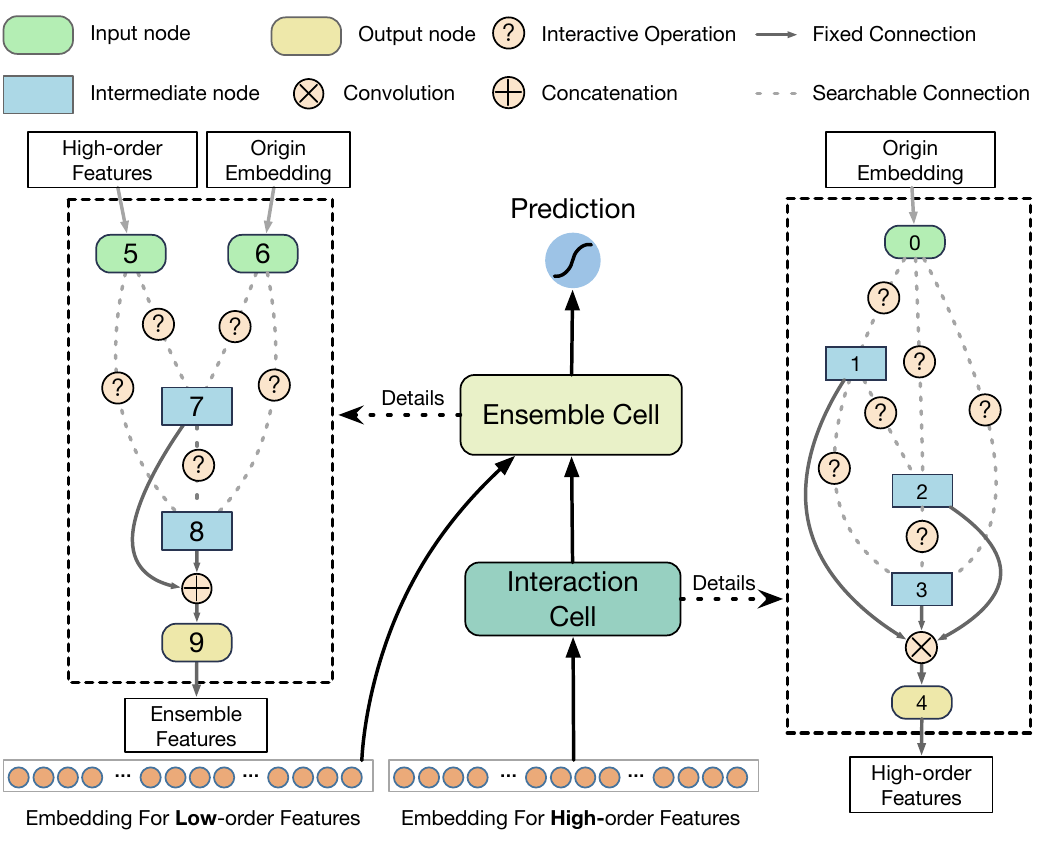}
  \caption{AutoPI decomposes the searchable part into an interaction cell and an ensemble cell. Each cell is represented as a DAG with several nodes, where each dashed line defines a set of candidate interactive operators.}
  \label{fig:cell}
\end{figure}
We first introduce two types of computational cells in this section.
As shown in Figure \ref{fig:cell}, each cell is a DAG consisting of an ordered sequence of nodes, including the input node(s), $N$ intermediate nodes and the output node, where $N$ is a hyperparameter, indicating the maximum order in interaction cell and the number of towers in ensemble cell.
Each node $\textbf{X}^{(i)}\in\mathbb{R}^{m\times k}$ is a latent representation (i.e., a feature matrix), 
and each directed edge $(i, j)$ is associated with an interactive operator $o^{(i, j)}$ that transforms $\textbf{X}^{(i)}$. 
And each intermediate node is computed based on all of its predecessors 
\begin{align}
    \textbf{X}^{(j)}=\sum_{i<j}o^{(i,j)}(\textbf{X}^{(i)})\label{eq:e3}.
\end{align}
The task of learning the cell, therefore, reduces to choosing the best operators for its edges.

We have two types of cells to serve two main functions: 
(1) \textit{the interaction cell formulates the higher-order feature interactions}. 
Specifically, the interaction cell has an input node (node $0$) that is defined 
as the input embedding produced from one of the dual-embedding layers, three intermediate nodes (node $1,2,3$) 
representing the intermediate cross features,
and an output node (node $4$) that fuses the feature matrices of all intermediate nodes via a combination operation (i.e., convolution).
(2) \textit{The ensemble cell formulates the ensemble of lower-order and higher-order interactions}. Different from the interaction cell, the ensemble cell has two input nodes, 
where node $5$ is the matrix of the higher-order cross features generated from the interaction cell, and node $6$ is the input embedding matrix produced by another of dual-embedding layers
Note that the intermediate nodes (node $7, 8$) in the ensemble cell function as the output from two towers. The output node $9$ is a concatenation of all intermediate nodes. Since the number of intermediate nodes is configurable, the number of towers in the ensemble cell can be adapted to the scenarios.
It is notable that although the cells are separate DAGs, instead of optimizing the cells sequentially, a joint optimization can also be applied.

In summary, we explicitly search for higher-order and lower-order interactions, and integrate them by a multi-tower structure.
According to our experiments, sharing embedding layers between deep and shallow models leads to a bad performance. Therefore, We adopt the dual-embedding layers to avoid the gradient interference between different cells \cite{cheng2020adaptive}.\\
\subsubsection{Operation Space.}
\label{sec:op}
To search for various interactions, diverse interactive operators should be included in our search space.
We extract and modulize interactive operators with the following considerations \cite{song2020towards}:\\
$\bullet$ \textbf{Functionality:} Operators in our search space should complement and accommodate each other for applications in extensive scenarios. For example, complex operators (e.g., SLP, convolution layer, etc.) have strong learning ability, modeling more complex cross features. In contrast, simple operators (e.g., LR, SENET Layer, etc.) can learn general representation with a low risk of overfitting.\\
$\bullet$ \textbf{Operator granularity:} The fine-grained operators (e.g., FM, SLP, etc.) have substantial combinational flexibility than coarse-grained (e.g., AFM, MLP, etc.). Because the NAS method is sensitive to computational overhead, efficient operation space requires operators with lower time-space complexity. Thus fine-grained operators are preferable.\\
$\bullet$ \textbf{Dimension alignment:} In our computation graph, we need to maintain the dimension of the output feature $o(\textbf{X})\in\mathbb{R}^{m \times k}$ as the same as the input feature $\textbf{X}\in\mathbb{R}^{ m \times k}$.

According to the above principles, we extract and modify the operators from representative NAS and CTR prediction literature, such as LR, inner product, hadamard product, outer product, cross layer \cite{wang2017deep}, none \cite{liu2018darts}, skip-connection, self-attention \cite{song2019autoint}, FM \cite{rendle2012factorization}, logarithmic transformation layer \cite{cheng2020adaptive}, SENET layer \cite{huang2019fibinet}, 2d convolution \cite{liu2019feature}, 1d convolution, etc. 
After extensive experiments and careful comparison, we finally determine the following operators to form the operation space:
\begin{itemize}
    \item \textit{Skip-connection} : takes an input feature matrix $\textbf{X}\in\mathbb{R}^{m\times k}$ and outputs the same feature matrix $\textbf{X}^{\prime}=\textbf{X}\in\mathbb{R}^{m\times k}$, which is an identity mapping and does not increase the order of interactions.
    \item \textit{SENET Layer} \cite{huang2019fibinet}: takes an input feature matrix $\textbf{X}\in\mathbb{R}^{m\times k}$ and produces a weight vector $\textbf{a} = [a_1, a_2, \dots, a_m]$, and then rescales $\textbf{X}$ with vector $\textbf{a}$ to obtain a weighted feature matrix $\textbf{X}^{\prime}=[a_1\cdot \textbf{x}_1, a_2\cdot \textbf{x}_2, \dots, a_m\cdot \textbf{x}_m]\in\mathbb{R}^{m\times k}$.
    \item \textit{Self-attention} \cite{song2019autoint}: takes an input feature matrix $\textbf{X}\in\mathbb{R}^{m\times k}$ and 
    outputs cross features $\textbf{X}^{\prime}\in\mathbb{R}^{m\times k}$ via the key-value attention mechanism.
    \item \textit{FM} : takes an input feature matrix $\textbf{X}\in\mathbb{R}^{m\times k}$ 
    and produces an inner product vector $\textbf{p}=\{(\textbf{x}_j\cdot \textbf{x}_j)\}_{(i,j)\in R_x}$, 
    where $R_x=\{(i,j)\}_{i\in\{1,\dots,m\}, j\in\{1,\dots,m\}, i<j}$, 
    and then outputs the feature matrix $\textbf{X}^{\prime}=\textbf{p}\cdot \textbf{W}\in \mathbb{R}^{m\times k}$ 
    via a linear transformation for dimension alignment.
    \item \textit{Single-layer Perceptron} : takes an input flattened feature vector $\textbf{x}\in\mathbb{R}^{mk}$ 
    and outputs a feature matrix $\textbf{X}^{\prime}=\textbf{x} \cdot \textbf{W}\in \mathbb{R}^{m\times k}$ 
    via a linear transformation.
    \item \textit{1d Convolution} : takes an input feature matrix $\textbf{X}\in\mathbb{R}^{m\times k}$ and 
    outputs a feature matrix $\textbf{X}^{\prime}\in\mathbb{R}^{m\times k}$ via 
    convolving with $m$ kernel matrices $\{\textbf{C}^{(i)}\}_{i\in\{1,\dots,m\}}\in\mathbb{R}^{m\times 1\times 1}$.
\end{itemize}
Note that each operator is followed by a nonlinear activation function (e.g. $ReLU(\cdot)$), excluding \textit{Skip-connection}. 

\subsection{Search Strategy}
\label{sec:st}

The gradient-based search strategy is adopted in our method for computational efficiency.
We first introduce architecture parameters for continuous relaxation on discrete search space. By doing so, 
the architecture can be optimized by the gradient descent method.
\subsubsection{Continuous Relaxation.}
\label{sec:cr}
Let $\mathcal{O}$ be a set of candidate interactive operators, where each operator is denoted by a function $o(\cdot)$ to be applied to $\textbf{X}^{(i)}$. 
Within a cell, the goal is to choose the best operator from $\mathcal{O}$ to connect each pair of nodes. 
Let a pair of nodes be $(i,j)$, where $i<j$, 
the key of continuous relaxation is to convert the combinatorial optimization problem to find the best weights of all operators between $(i,j)$ after formulating the information propagated from $i$ to $j$ as a weighted sum over $|\mathcal{O}|$ operators, namely,
\begin{align}
    f^{(i, j)}(\textbf{X}^{(i)})=\sum_{o \in \mathcal{O}} \frac{\exp \left(\alpha_{o}^{(i, j)}/\tau\right)}{\sum_{o^{\prime} \in \mathcal{O}} \exp \left(\alpha_{o^{\prime}}^{(i, j)}/\tau\right)} o(\textbf{X}^{(i)})
\label{eq:ops}
\end{align}
where the weights of operators are parameterized by a vector $\alpha^{(i,j)}$ with dimension $|\mathcal{O}|$, and 
$\tau > 0$ is a parameter called temperature for controlling the importance discrepancy among all operators. 
Besides operation-level parameters $\alpha$, we also introduce edge-level parameters $\beta$ for choosing important node pairs in an interaction cell. 
And the Equation \ref{eq:e3} can be rewrote as:
\begin{align}
    \textbf{X}^{(j)}=\sum_{i<j} \frac{\exp \left(\beta^{(i, j)}/\tau\right)}{\sum_{i^{\prime}<j} \exp \left(\beta^{(i^{\prime}, j)}/\tau\right)} \cdot f^{(i, j)}(\textbf{X}^{(i)}).
\label{eq:edge}
\end{align}
As a result, we omit the \textit{none} operation widely used in most gradient-based NAS methods \cite{liu2018darts,xu2019pc,chu2020noisy,li2020pd,liang2019darts+}, and normalize $\textbf{X}^{(j)}$ for numerical stability. The task of architecture search then reduces to learning two set of continuous variables $\alpha=\{\alpha^{(i,j)}\}_{i<j}$, $\beta=\{\beta^{(i,j)}\}_{i<j}$.
\subsubsection{Bi-level Optimization.}
\label{sec:opt}
Following \cite{liu2018darts}, we use the Bi-level algorithm to alternatively learn the architecture parameters $\alpha, \beta$ and the weights $w$ of all operators in the architecture.
Specifically, the goal for searching powerful interactions is to find $\alpha^*, \beta^*$ that minimizes the validation loss $\mathcal{L}_{val}(w^*, \alpha, \beta)$, 
where the weights $w^*$ are obtained by minimizing the training loss $w^*=\operatorname{argmin}_w\mathcal{L}_{train}(w, \alpha, \beta)$.
This implies a bilevel optimization problem with $\alpha, \beta$ as the upper-level variable and $w$ as the lower-level variable.
Evaluating the architecture gradient exactly can be prohibitive due to the expensive inner optimization.
Therefore, the one-step approximation \cite{liu2018darts} is used and the approximate architecture gradient yields

\begin{align}
\nabla_{\alpha,\beta} \mathcal{L}_{v a l}\left(w^{\prime}, \alpha,\beta\right)-\xi \nabla_{\alpha,\beta, w}^{2} \mathcal{L}_{\text {train}}(w, \alpha,\beta) \nabla_{w^{\prime}} \mathcal{L}_{\text {val}}\left(w^{\prime}, \alpha,\beta\right)
\end{align}
where $w^{\prime}=w-\xi\nabla_{w}\mathcal{L}_{train}(w,\alpha,\beta)$ denotes the weights for one-step forward model. 
We use the first-order approximation following \cite{xu2019pc}, w.r.t $\xi=0$, to reduce the computational complexity in the search process.
\subsubsection{Batch Normalization.}
\label{sec:bn}
In Equation \ref{eq:ops}, the node $j$ is the weighted sum of outputs of all operators in the operation space given $i$ as the input. 
The contribution of a feature interaction $o(\textbf{X}^{(i)})$ can be measured by 
$\mathbf{softmax}(\alpha_{o}^{(i,j)})$.
However, two problems may arise in fairness: (1) Our operation space contains both complicated and simple operators. In Bi-level optimization, the simple operators generally converge faster than complex operators, and then the stability of converged simple operators yield continuously better performance than unstable complex operators, which promotes larger weights $\mathbf{softmax}(\alpha_{o}^{(i,j)})$ on simple operators. 
Since each edge reserves only one operator with the largest weight, more simple operators are prone to be selected. 
Therefore, the optimization on architecture parameters $\alpha, \beta$ should consider the comparison fairness of converged and under-converged operators.
(2)
Since the range of $o(\textbf{X}^{(i)})$ varies among different operators
and $o(\textbf{X}^{(i)})$ is jointly optimized with $\mathbf{softmax}(\alpha_{o}^{(i,j)})$,
the coupling of their scale will lead to that $\mathbf{softmax}(\alpha_{o}^{(i,j)})$ can hardly represent the relative importance of $o(\textbf{X}^{(i)})$ \cite{liu2020autofis}.

It is well known that BN \cite{ioffe2015batch} can effectively solve the above problems by moving the output of each operator to a standard normal distribution, which is expressed as
\begin{equation}
BN(o(\textbf{X}))=\frac{o(\textbf{X})-\mu_{\mathcal{B}}\left(o(\textbf{X})\right)}{\sqrt{\sigma_{\mathcal{B}}^{2}\left(o(\textbf{X})\right)+\epsilon}}
\end{equation}
where $\mu_\mathcal{B}$ and $\sigma_\mathcal{B}$ are the mean, 
standard deviation vector of $o(\textbf{X})$ on the field dimension in mini-batch $\mathcal{B}$ and $\epsilon$ is a constant for numerical stability.
Note that there is no need to optimize to distribution after normalization. Therefore, we set the scale and shift parameters in BN to be 1 and 0, respectively.

\subsection{Performance Evaluation}
\label{sec:pe}
The objective of our method is to find powerful interactions that can achieve high predictive performance on a target data distribution.
We obtain architecture parameters with the best performance on the validation set during the search process, and then convert the continuous architecture encoding to a discrete architecture. The discrete architecture will be trained from scratch.
However, a well-known problem in weight-sharing NAS methods is the performance gap \cite{xie2020weight} between the continuous architecture encoding and the derived discrete architecture, which is called discrete loss.
In this section, we first introduce the discretization technique, and then illustrate two optional training techniques for reducing the performance gap.

\begin{table}[]
\caption{Statistics of the datasets.}
\label{tab:dataset}
\begin{tabular}{@{}cccc@{}}
\toprule
Dataset & \#instance & \#fields & \#features \\ \midrule
Criteo & 45,840,617 & 39 & 2,086,936 \\
Avazu & 40,428,967 & 22 & 1,544,250 \\
Movielens & 2,006,859 & 3 & 90,445 \\
Frappe & 288,609 & 10 & 5382 \\
S1-S5 & $\sim$25,000,000*5 & 28*5 & $\sim$600,000*5 \\ \bottomrule
\end{tabular}
\end{table}

\subsubsection{Discretization.}
\label{sec:dis}
For each node in the converged computational graph, we retain the top-$k$ strongest operators (from distinct nodes) among all candidate operators collected from all previous nodes. The strength of an operator is defined as $\frac{\text{exp}(\alpha_{o}^{(i,j)})}{\sum_{o^{\prime}\in\mathcal{O}}\text{exp}(\alpha_{o^{\prime}}^{(i,j)})}\cdot \frac{\text{exp}(\beta^{(i,j)})}{\sum_{k<j}\text{exp}(\beta^{(k,j)})}$. We use $k=2$ for interaction cell and ensemble cell following existing works on CV domain \cite{liu2018darts,xu2019pc}.\\
\subsubsection{Reducing Discrete Loss.}
\label{sec:rdl}
Two training techniques can be optionally utilized to reduce the performance gap in our method.\\
(1) \textit{Temperature Anneal}.
\cite{liu2018darts} alleviate the performance gap via annealing the temperature parameter in softmax.
By doing so, in each node pair, all operators have nearly the same strength defined above in order to train the weights $w$ in the early search stage. 
As the temperature decreases, operators and edges begin to compete with each other to increase the discrepancy.  
We anneal the softmax temperature on every training epoch as follow formulation:
\begin{align}
\label{eq:anneal}
\tau=\frac{1+\text{log}(T)}{1+\text{log}(t)}
\end{align}
where $T$ is the number of training epochs,  and $t\in[1,\dots,T]$ is an index of the current epoch.\\
(2) \textit{Noisy Skip-connection}.
\cite{chu2020noisy} proved that \textit{Skip-connection} usually overperform other operators with softmax, 
resulting in a degenerate model where the skip connections contribute less. 
After injecting noise into the \textit{Skip-connection} operator, the bias towards \textit{Skip-connection} will be weakened. 
In this paper, we inject the Gaussian noise $\tilde{\textbf{X}} \sim \mathcal{N}(\mu, \sigma)$ into our \textit{Skip-connection} operator, i.e., $o_{\text{skip}}({\textbf{X}})={\textbf{X}}+\tilde{{\textbf{X}}}$, 
where $\mu=0$ and $\sigma=\lambda\cdot std({\textbf{X}})$. Note that $\lambda$ is a small positive 
coefficient.
\section{Experiments and Results}
\label{sec:exp}
To comprehensively evaluate our method, we design plenty of experiments to answer the following research questions: \\
\textbf{RQ1:} How do the architectures identified via AutoPI perform compared with the SOTA hand-crafted models and available NAS algorithms on datasets with different scales?\\
\textbf{RQ2:} How efficient is AutoPI compared with the SOTA hand-crafted models and available NAS algorithms?\\
\textbf{RQ3:} How effective are training techniques in alleviating the performance gap between the networks before and after discretization?\\
\textbf{RQ4:} How to understand the computational graph and operation space in our search space?\\
\subsection{Experiment setup}
\begin{table*}[]
\caption{Effectiveness Comparison of Different Algorithms on Public Benchmarks.}
\label{tab:pub}
\resizebox{\textwidth}{!}{
\begin{tabular}{@{}cccccccccccc@{}}
\toprule
\multirow{2}{*}{Model Class} & \multirow{2}{*}{Model} & \multicolumn{2}{c}{Criteo} & \multicolumn{2}{c}{Avazu} & \multicolumn{2}{c}{Movielens} & \multicolumn{2}{c}{Frappe} & \multirow{2}{*}{$\Delta_{AUC}\uparrow$} & \multirow{2}{*}{$\Delta_{Logloss}\downarrow$} \\ \cmidrule(lr){3-10}
 &  & AUC & Logloss & AUC & Logloss & AUC & Logloss & AUC & Logloss &  &  \\ \midrule
\multicolumn{1}{c|}{First-Order} & \multicolumn{1}{c|}{LR} & 0.7858 & 0.4636 & 0.7313 & 0.4065 & 0.9215 & 0.3080 & 0.9329 & \multicolumn{1}{c|}{0.2860} & -3.01\% & +0.0535 \\ \midrule
\multicolumn{1}{c|}{\multirow{2}{*}{Second-Order}} & \multicolumn{1}{c|}{FM} & 0.7933 & 0.4574 & 0.7496 & 0.3740 & 0.9388 & 0.2797 & 0.9641 & \multicolumn{1}{c|}{0.2143} & -1.16\% & +0.0188 \\
\multicolumn{1}{c|}{} & \multicolumn{1}{c|}{AFM} & 0.7953 & 0.4554 & 0.7454 & 0.3766 & 0.9295 & 0.2836 & 0.9639 & \multicolumn{1}{c|}{0.2294} & -1.45\% & +0.0237 \\ \midrule
\multicolumn{1}{c|}{\multirow{7}{*}{High-Order}} & \multicolumn{1}{c|}{CrossNet} & 0.7915 & 0.4585 & 0.7498 & 0.3756 & 0.9323 & 0.2929 & 0.9393 & \multicolumn{1}{c|}{0.2835} & -1.98\% & +0.0400 \\
\multicolumn{1}{c|}{} & \multicolumn{1}{c|}{HOFM} & 0.7960 & 0.4551 & 0.7516 & 0.3756 & 0.9410 & 0.3088 & 0.9709 & \multicolumn{1}{c|}{0.2141} & -0.81\% & +0.0259 \\
\multicolumn{1}{c|}{} & \multicolumn{1}{c|}{NFM} & 0.7968 & 0.4537 & 0.7531 & 0.3761 & 0.9441 & 0.3004 & 0.9727 & \multicolumn{1}{c|}{0.2079} & -0.63\% & +0.0220 \\
\multicolumn{1}{c|}{} & \multicolumn{1}{c|}{PNN} & 0.8026 & 0.4509 & 0.7526 & 0.3737 & 0.9469 & 0.2792 & 0.9735 & \multicolumn{1}{c|}{0.2012} & -0.41\% & +0.0137 \\
\multicolumn{1}{c|}{} & \multicolumn{1}{c|}{CIN} & 0.8042 & 0.4472 & 0.7533 & 0.3756 & 0.9494 & 0.2600 & 0.9704 & \multicolumn{1}{c|}{0.2342} & -0.37\% & +0.0170 \\
\multicolumn{1}{c|}{} & \multicolumn{1}{c|}{AutoInt} & 0.8062 & 0.4456 & 0.7465 & 0.3790 & 0.9418 & 0.2762 & 0.9701 & \multicolumn{1}{c|}{0.2034} & -0.69\% & +0.0135 \\
\multicolumn{1}{c|}{} & \multicolumn{1}{c|}{AFN} & 0.8061 & 0.4458 & 0.7512 & 0.3731 & 0.9477 & 0.2753 & 0.9759 & \multicolumn{1}{c|}{0.1784} & -0.28\% & +0.0056 \\ \midrule
\multicolumn{1}{c|}{\multirow{6}{*}{Ensembled}} & \multicolumn{1}{c|}{Deep\&Cross} & 0.8059 & 0.4463 & 0.7550 & 0.3721 & 0.9419 & 0.2791 & 0.9402 & \multicolumn{1}{c|}{0.2808} & -1.23\% & +0.0320 \\
\multicolumn{1}{c|}{} & \multicolumn{1}{c|}{Wide\&Deep} & 0.8062 & 0.4453 & 0.7529 & 0.3744 & 0.9381 & 0.3310 & 0.9728 & \multicolumn{1}{c|}{0.2038} & -0.55\% & +0.0261 \\
\multicolumn{1}{c|}{} & \multicolumn{1}{c|}{DeepFM} & 0.8025 & 0.4501 & 0.7535 & 0.3742 & 0.9424 & 0.3131 & 0.9719 & \multicolumn{1}{c|}{0.2108} & -0.54\% & +0.0245 \\
\multicolumn{1}{c|}{} & \multicolumn{1}{c|}{xDeepFM} & 0.8070 & 0.4443 & 0.7535 & 0.3737 & 0.9448 & 0.2717 & 0.9738 & \multicolumn{1}{c|}{0.2098} & -0.32\% & +0.0123 \\
\multicolumn{1}{c|}{} & \multicolumn{1}{c|}{AutoInt+} & 0.8075 & 0.4438 & 0.7477 & 0.3776 & 0.9417 & 0.2764 & 0.9661 & \multicolumn{1}{c|}{0.2102} & -0.73\% & +0.0145 \\
\multicolumn{1}{c|}{} & \multicolumn{1}{c|}{AFN+} & 0.8083 & 0.4437 & 0.7555 & 0.3718 & 0.9500 & 0.2585 & 0.9783 & \multicolumn{1}{c|}{0.1762} & - & - \\ \midrule
\multicolumn{1}{c|}{\multirow{5}{*}{NAS}} & \multicolumn{1}{c|}{AutoRec-R} & 0.8104 & 0.4412 & 0.7499 & 0.3746 & 0.9510 & 0.2573 & 0.9775 & \multicolumn{1}{c|}{0.1688} & -0.08\% & -0.0021 \\
\multicolumn{1}{c|}{} & \multicolumn{1}{c|}{AutoRec-G} & 0.8100 & 0.4417 & 0.7467 & 0.3756 & 0.9496 & 0.2579 & 0.9771 & \multicolumn{1}{c|}{0.1778} & -0.22\% & +0.0007 \\
\multicolumn{1}{c|}{} & \multicolumn{1}{c|}{AutoRec-B} & 0.8097 & 0.4418 & 0.7488 & 0.3766 & 0.9539 & \textbf{0.2443} & 0.9783 & \multicolumn{1}{c|}{0.1610} & -0.03\% & -0.0066 \\ \cmidrule(l){2-12} 
\multicolumn{1}{c|}{} & \multicolumn{1}{c|}{AutoPI-R} & \textbf{0.8105} & \textbf{0.4410} & 0.7598 & 0.3730 & 0.9555 & 0.2598 & 0.9820 & \multicolumn{1}{c|}{0.1487} & +0.39\% & -0.0069 \\
\multicolumn{1}{c|}{} & \multicolumn{1}{c|}{\textbf{AutoPI-D}} & 0.8102 & 0.4420 & \textbf{0.7605} & \textbf{0.3691} & \textbf{0.9572} & 0.2660 & \textbf{0.9827} & \multicolumn{1}{c|}{\textbf{0.1426}} & \textbf{+0.46\%} & \textbf{-0.0076} \\ \bottomrule
\end{tabular}
}
\end{table*}
\textbf{Datasets.} We conduct experiments on four public datasets with different statistical characteristics and five private commercial datasets generated from different E-commerce scenarios.
The setting of public datasets follows previous works \cite{cheng2020adaptive,lian2018xdeepfm,he2017neural}, including 
Criteo$\footnote{http://labs.criteo.com/2014/02/kaggle-display-advertising- challenge-dataset/}$, Avazu$\footnote{https://www.kaggle.com/c/avazu-ctr-prediction}$, Movielens$\footnote{https://grouplens.org/datasets/movielens/}$ and Frappe$\footnote{http://baltrunas.info/research-menu/frappe}$. We randomly split the instances by 8:1:1 for training, validation and testing, respectively.
The commercial datasets contain users' browsing and click records generated from real-world traffic logs on an E-commerce website. It is collected from five different scenes (S1-S5) in a discount activity. Each record has its corresponding user, item, context features, and a label indicating a click or not. 
Data of 3 days are used for training, and the following two days are used for validation and testing, respectively.
The statistics of all datasets is shown in Table \ref{tab:dataset}.\\
\textbf{Evaluation metrics.} We adopt two metrics for performance evaluation: AUC (Area Under the ROC curve) and Logloss (cross-entropy). Due to a large number of datasets, $\Delta_{AUC}$ and $\Delta_{Logloss}$ are also calculated to indicate averaged performance gain compared to a given benchmark over a group of datasets, which help to reflect the generalization ability of the algorithms.
Note that an increase in AUC or decrease in Logloss at .001-level is known to be a significant improvement for the CTR prediction task \cite{cheng2016wide,song2019autoint,cheng2020adaptive}.\\
\textbf{Comparison methods.} We compare \textbf{AutoPI-D} (\textbf{AutoPI} with Bi-level optimization in \textbf{D}ARTS) with five classes of the existing approaches: 
(i) first-order approaches that is a weighted sum of raw features, including \textbf{LR}; 
(ii) FM-based methods that consider second-order cross features, including \textbf{FM} \cite{rendle2012factorization} and \textbf{AFM} \cite{xiao2017attentional}; 
(iii) advanced approaches that model higher-order feature interactions, including \textbf{CrossNet} \cite{wang2017deep}, \textbf{HOFM} \cite{blondel2016higher}, \textbf{NFM} \cite{he2017neural}, \textbf{PNN} \cite{qu2018product}, \textbf{CIN} \cite{lian2018xdeepfm}, \textbf{AutoInt} \cite{song2019autoint} and \textbf{AFN} \cite{cheng2020adaptive}; 
(iv) ensemble models that are integrated with a tower of DNN, including \textbf{Deep\&Cross} \cite{wang2017deep}, \textbf{Wide\&Deep} \cite{cheng2016wide}, \textbf{DeepFM} \cite{guo2017deepfm}, \textbf{xDeepFM} \cite{lian2018xdeepfm}, \textbf{AutoInt+} \cite{song2019autoint} and \textbf{AFN+} \cite{cheng2020adaptive}.
(v) NAS algorithms that automatically search the effective architecture for a target dataset, including \textbf{AutoRec} \cite{wang2020autorec} and \textbf{AutoPI-R} (AutoPI with random search strategy). Note that no available codes have been found for other NAS methods, such as AutoCTR. We select the open-sourced \textbf{AutoRec} as our baseline. And we denote \textbf{AutoRec-R}, \textbf{AutoRec-G}, \textbf{AutoRec-B} as the random, greedy, Bayesian version of \textbf{AutoRec}, respectively.\\
\textbf{Implementation details.} We implement AutoPI using 
Pytorch$\footnote{The code is available here: \url{https://github.com/thu-media/AutoPI}.}$.
Both interaction and ensemble cells consist of $N=4$ intermediate nodes. We set the embedding size as $k=10$ and adopt temperature anneal during the search process.
Half of the training set is used for optimizing weights $w$ in operators, and the other half is for architecture parameters $\alpha, \beta$.
The maximum number of training epochs for Criteo, Avazu, Commerical Dataset, Frappe, Movielens are 5, 5, 5, 50 and 100, respectively, and the corresponding batch size is 4096.
We use momentum SGD to optimize the weights $w$ and use Adam as the optimizer for architecture parameters $\alpha, \beta$, with the configuration of \cite{xu2019pc}. To avoid overfitting, we perform early-stopping according to the Logloss on the validation set. Furthermore, we implement hand-crafted models according to the details of \cite{cheng2020adaptive}, 
where we use the same neural network structure (i.e., 3-layers MLP, 400-400-400) for all approaches that involve DNN. We implement the NAS baselines by following \cite{wang2020autorec}. For each empirical result, we run the experiments with different random seeds for three times and report the average value.
\begin{table*}[]
\caption{Effectiveness Comparison of Different Algorithms on Real Scenarios.}
\label{tab:dacu}
\resizebox{\textwidth}{!}{
\begin{tabular}{@{}ccccccccccccc@{}}
\toprule
\multirow{2}{*}{Model} & \multicolumn{2}{c}{S1} & \multicolumn{2}{c}{S2} & \multicolumn{2}{c}{S3} & \multicolumn{2}{c}{S4} & \multicolumn{2}{c}{S5} & \multirow{2}{*}{\begin{tabular}[c]{@{}c@{}}$\Delta_{AUC}$\\ $\uparrow$\end{tabular}} & \multirow{2}{*}{\begin{tabular}[c]{@{}c@{}}$\Delta_{Logloss}$\\ $\downarrow$\end{tabular}} \\ \cmidrule(lr){2-11}
 & AUC & Logloss & AUC & Logloss & AUC & Logloss & AUC & Logloss & AUC & Logloss &  &  \\ \midrule
\multicolumn{1}{c|}{xDeepFM} & 0.7973 & 0.1678 & 0.7179 & 0.3319 & 0.7481 & 0.1462 & 0.7457 & 0.1449 & 0.8206 & 0.2970 & -0.67\% & +0.0030 \\
\multicolumn{1}{c|}{AutoInt+} & 0.8033 & 0.1639 & 0.7173 & 0.3319 &  0.7606 & 0.1396 & 0.7400 & 0.1459 &0.8214 & 0.2949 & -0.41\%  & +0.0006 \\
\multicolumn{1}{c|}{AFN+} & 0.8084 & \textbf{0.1618} & 0.7263 & 0.3291 & 0.7610 & 0.1414 & 0.7464 & 0.1448 & 0.8208 & 0.2954 & - & - \\ \midrule
\multicolumn{1}{c|}{AutoRec-R} & 0.8112 & 0.1634 & 0.7230 & 0.3300 & 0.7637 & 0.1413 & 0.7502 & 0.1445 & 0.8248 & 0.2930 & +0.20\% & -0.0001 \\
\multicolumn{1}{c|}{AutoRec-G} & 0.8082 & 0.1637 & 0.7214 & 0.3307 & 0.7638 & 0.1414 & 0.7509 & 0.1445 & 0.8248 & \textbf{0.2929} & +0.12\% & +0.0001 \\
\multicolumn{1}{c|}{AutoRec-B} & 0.8072 & 0.1636 & 0.7235 & 0.3298 & 0.7630 & 0.1407 & 0.7512 & 0.1442 & 0.8250 & 0.2930 & +0.14\% & -0.0001 \\ \midrule
\multicolumn{1}{c|}{AutoPI-R} & 0.8128 & 0.1647 & 0.7308 & 0.3302 & 0.7656 & 0.1392 & 0.7545 & 0.1434 & 0.8238 & 0.2948 & +0.50\% & -0.0001 \\
\multicolumn{1}{c|}{\textbf{AutoPI-D}} & \textbf{0.8142} & 0.1638 & \textbf{0.7316} & \textbf{0.3287} & \textbf{0.7678} & \textbf{0.1387} & \textbf{0.7590} & \textbf{0.1431} & \textbf{0.8256} & 0.2970 & \textbf{+0.70\%} & \textbf{-0.0002} \\ \bottomrule
\end{tabular}
}
\end{table*}
\begin{table*}[]
\caption{Comparison Of Architecture Efficiency Among SOTA Methods (Excluding Embedding Layers Of The Same Size).}
\label{tab:cplx}
\resizebox{\textwidth}{!}{
\begin{tabular}{@{}ccccccccccccc@{}}
\toprule
\multirow{2}{*}{Model} & \multicolumn{4}{c}{\# Params (Million)} & \multicolumn{4}{c}{Flops (Million)} & \multicolumn{4}{c}{Search cost (GPU Hours)} \\ \cmidrule(l){2-13} 
 & Criteo & Avazu & Mvl & Frappe & Criteo & Avazu & Mvl & Frappe & Criteo & Avazu & Mvl & Frappe \\ \midrule
\multicolumn{1}{c|}{xDeepFM} & 7.33 & 4.13 & 0.82 & \multicolumn{1}{c|}{2.01} & 31.24 & 11.54 & 1.16 & \multicolumn{1}{c|}{3.55} & - & - & - & - \\
\multicolumn{1}{c|}{AutoInt+} & 0.49 & 0.42 & 0.34 &  \multicolumn{1}{c|}{0.37} & 27.79 & 6.75 & 0.89 & \multicolumn{1}{c|}{1.86} & - & - & - & - \\
\multicolumn{1}{c|}{AFN+} & 6.86 & 5.56 & 3.86 & \multicolumn{1}{c|}{3.09} & 14.18 & 11.32 & 7.73 & \multicolumn{1}{c|}{6.22} & - & - & - & - \\ \midrule
\multicolumn{1}{c|}{AutoRec-R} & 1.47 & 0.01 & 0.04 & \multicolumn{1}{c|}{0.03} & 2.82 & 0.01 & 0.07 & \multicolumn{1}{c|}{0.05} & 47.50 & 22.20 & 1.70 & 1.30 \\
\multicolumn{1}{c|}{AutoRec-G} & 0.58 & 0.01 & 0.30 & \multicolumn{1}{c|}{0.04} & 1.17 & 0.01 & 0.61 & \multicolumn{1}{c|}{0.07} & 45.70 & 18.80 & 2.10 & 1.60 \\
\multicolumn{1}{c|}{AutoRec-B} & 0.85 & 0.01 & 1.21 & \multicolumn{1}{c|}{0.02} & 1.70 & 0.02 & 2.50 & \multicolumn{1}{c|}{0.05} & 41.10 & 20.20 & 20.50 & 2.40 \\ \midrule
\multicolumn{1}{c|}{AutoPI-R} & 1.60 & 0.14 & 0.01 & \multicolumn{1}{c|}{0.02} & 3.46 & 0.32 & 0.02 & \multicolumn{1}{c|}{0.07} & 50.50 & 19.40 & 2.00 & 1.20 \\
\multicolumn{1}{c|}{AutoPI-D} & 1.18 & 0.10 & 0.01 & \multicolumn{1}{c|}{0.05} & 2.51 & 0.25 & 0.02 & \multicolumn{1}{c|}{0.12} & \textbf{4.20} & \textbf{2.20} & \textbf{0.40} & \textbf{0.10} \\ \bottomrule
\end{tabular}
}
\end{table*}
\subsection{Overall Performance (RQ1)}
\label{sec:rq1}
We set the powerful hand-crafted model, AFN+, as the benchmark for calculating $\Delta_{AUC}$ and $\Delta_{Logloss}$ on all datasets.\\
\textbf{Public Datasets.}
As illustrated in Table \ref{tab:dataset}, the public datasets have distinct heterogeneous feature spaces, 
w.r.t, different number of instances, fields and features.
Table \ref{tab:pub} summarizes the performance of all methods on four public datasets.

We first present three important observations when comparing various hand-crafted models. 
First, the methods that utilize higher-order feature interactions generally outperform those based on lower-order cross features, which means higher-order feature interactions are informative. 
Second, AFN and AutoInt consistently outperform FMs and HOFMs on all public datasets, which verifies that learning adaptive-order cross features can bring better predictive performance than modeling fixed-order feature interactions.
Third, the ensemble models have a significant performance improvement than individual models, which demonstrates that combining different types of interactions is more effective than a single one.

Moreover, the NAS baseline, AutoRec, produces the competitive performance comparing with the AFN+, and outperforms all other hand-crafted models. This illustrates that using NAS can successfully find effective feature interactions and alleviate laborious and tedious architecture engineering.
The search space in AutoRec supports the ensemble of at most 6 interactive blocks, where each block can be selected from one of the 5 types of configurable operators: MLP, FM, CrossNet, self-attention and element-wise interaction. Although the number of blocks and hyperparameters in the operators can be jointly optimized by different algorithms including the effective Bayesian method, the search space of AutoRec only incorporates limited efficient manual designs. AutoPI-R, exploring our search space by a simple search strategy, can search for more powerful interactions than AutoRec, which means our search space is more flexible and general than AutoRec.
We further compare our method (AutoPI-D) with AutoPI-R. AutoPI-D yields a better or competing performance than AutoPI-R over all public datasets. This demonstrates that the improved gradient-based search strategy is more effective than the random method.

\textbf{Real Scenarios.}
As illustrated in Table \ref{tab:dataset}, the commercial datasets have close statistical properties but diverse semantics.
We collect real-world traffic logs of five different scenes (S1-S5) from an online retailer platform. Each scene is a specific page on the E-commerce website to satisfy all kinds of shopping needs of users. The five scenes we collected contain food, makeup, etc. Each scene has different UI pages, target user groups, and data distribution. 
Due to page limitations, we select three representative hand-crafted models for real-scenario evaluation, including xDeepFM, AutoInt+ and AFN+.
The comparison on the commercial datasets of these algorithms is shown in Table \ref{tab:dacu}.
From the results, we observe that the NAS methods outperform all hand-crafted models over all commercial datasets. This illustrates that the NAS methods are more effective than hand-crafted models on extensive real scenarios.
Most importantly, our method increases the average AUC $\Delta_{AUC}$ by $0.70\%$, which is large than AutoRec by $0.50\%$, which further verify the generalization of our proposed AutoPI.
\subsection{Efficiency Analysis (RQ2)}
\label{sec:rq2}
\begin{table*}[]
\caption{Ablation study of interactive operators.}
\label{tab:ops-res}
\resizebox{\textwidth}{!}{
\begin{tabular}{@{}ccccccccccc@{}}
\toprule
\multirow{2}{*}{Omitted Operation} & \multicolumn{4}{c}{\textbf{Avazu}} & \multicolumn{4}{c}{Frappe} & \multirow{2}{*}{\begin{tabular}[c]{@{}c@{}}$\Delta_{AUC}$\\ $\uparrow$\end{tabular}} & \multirow{2}{*}{\begin{tabular}[c]{@{}c@{}}$\Delta_{Logloss}$\\ $\downarrow$\end{tabular}} \\ \cmidrule(lr){2-9}
 & AUC & Logloss & Params (M) & Flops (M) & AUC & Logloss & Params (M) & Flops (M) &  &  \\ \midrule
\multicolumn{1}{c|}{MLP} & 0.7580 & 0.3695 & 0.05 & 0.27 & 0.9790 & 0.1777 & 0.02 & \multicolumn{1}{c|}{0.07} & -0.62\% & +0.0166 \\
\multicolumn{1}{c|}{FM} & 0.7587 & \textbf{0.3695} & 0.10 & 0.24 & 0.9820 & \textbf{0.1399} & 0.05 & \multicolumn{1}{c|}{0.11} & -0.13\% & \textbf{-0.0024} \\
\multicolumn{1}{c|}{MLP\&FM} & 0.7594 & 0.3708 & \textbf{0.01} & \textbf{0.17} & 0.9685 & 0.2146 & \textbf{0.01} & \multicolumn{1}{c|}{\textbf{0.03}} & -0.77\% & +0.0357 \\
\multicolumn{1}{c|}{None} & \textbf{0.7605} & 0.3715 & 0.10 & 0.25 & \textbf{0.9827} & 0.1426 & 0.05 & \multicolumn{1}{c|}{0.12} & - & - \\ \bottomrule
\end{tabular}
}
\end{table*}
In real-world scenarios, both the effectiveness and efficiency of a method are critical. Many SOTA methods can achieve good accuracy but fail to be deployed in commercial recommender systems due to their computational inefficiency, especially the additional time-consuming search process in NAS methods.
In this section, we study the efficiency of AutoPI on public datasets using one NVIDIA P100 GPU.
We present the search time of the NAS algorithms and the complexity (parameters size and Flops) of the identified best architectures in Table \ref{tab:cplx}.
Benefiting from the gradient-based search strategy and weight-sharing mechanism, our method (AutoPI-D) requires $\sim \times10$ less search time than other NAS algorithms (AutoRec and AutoPI-D).
In addition, the complexity of the architectures identified via our method are close to AutoRec, but much lower than hand-crafted models \footnote{Although AutoInt+ has lower parameters than our method, the Flops of AutoInt+ is much larger than our method because of the key-value attention mechanism.}. 
Therefore, AutoPI can achieve better performance efficiently, which makes it easier to be deployed in industrial applications.

Furthermore, we present the ablation study of operators with large parameters in Table \ref{tab:ops-res} to reduce the complexity of our architectures.
In our operation space, MLP and FM have a large number of parameters, and we run four experiments on Avazu and Frappe with different configurations of operation space: (1) omitting MLP, (2) omitting FM, (3) omitting MLP and FM, (4) non-omitting. 
As illustrated in Table \ref{tab:ops-res}, the flexibility of AutoPI is validated. Therefore, by omitting operators with large parameters, the trade-off between performance and computational cost can be achieved.
\subsection{Training Techniques Study (RQ3)}
\begin{table}[]
\caption{Effectiveness Comparison of Training Techniques.}
\label{tab:train-res}
\resizebox{0.5\textwidth}{!}{
\setlength{\tabcolsep}{2mm}{
\begin{tabular}{@{}ccccccc@{}}
\toprule
\multirow{2}{*}{Techniques} & \multicolumn{2}{c}{Avazu} & \multicolumn{2}{c}{Frappe} & \multirow{2}{*}{\begin{tabular}[c]{@{}c@{}}$\Delta_{AUC}$\\ $\uparrow$\end{tabular}} & \multirow{2}{*}{\begin{tabular}[c]{@{}c@{}}$\Delta_{Logloss}$\\ $\downarrow$\end{tabular}} \\ \cmidrule(lr){2-5}
 & AUC & Logloss & AUC & Logloss &  &  \\ \midrule
\multicolumn{1}{c|}{Origin} & 0.7606 & 0.3688 & 0.9794 & \multicolumn{1}{c|}{0.1588} & - & - \\
\multicolumn{1}{c|}{Noisy} & \textbf{0.7610} & \textbf{0.3682} & 0.9809 & \multicolumn{1}{c|}{\textbf{0.1522}} & +0.10\% & \textbf{-0.0036} \\
\multicolumn{1}{c|}{Anneal} & 0.7605 & 0.3715 & \textbf{0.9827} & \multicolumn{1}{c|}{0.1553} & \textbf{+0.17\%} & -0.0004 \\
\multicolumn{1}{c|}{Noisy+Anneal} & 0.7605 & 0.3684 & 0.9804 & \multicolumn{1}{c|}{0.1556} & +0.09\% & -0.0018 \\ \bottomrule
\end{tabular}}
}
\end{table}
\begin{figure}[]
  \centering
  \includegraphics[width=\linewidth]{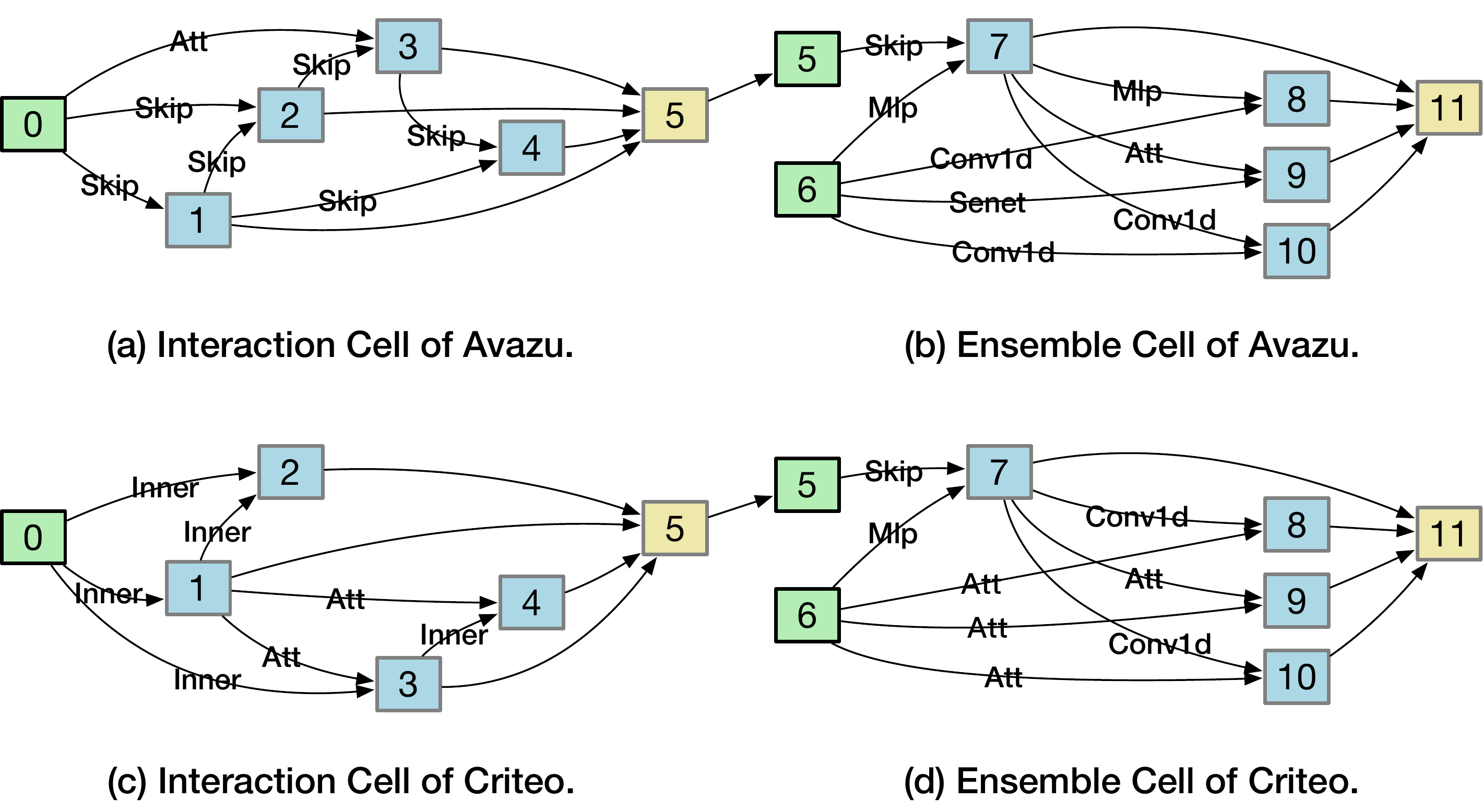}
  \vspace{-10pt}
  \caption{Two architectures found by AutoPI.}
  \vspace{-10pt}
  \label{fig:searched}
\end{figure}
In this section, we evaluate the effectiveness of two training techniques on Avazu and Frappe by ablation study and the results are illustrated in Table \ref{tab:train-res}. 
In detail, we run four experiments with different training techniques: 
(1) noisy skip-connection (Noisy): injecting the noise ($\lambda=0.03$) into the skip-connection with the fixed softmax temperature ($\tau=1$), 
(2) temperature anneal (Anneal): annealing the softmax temperature with the original skip-connection during the search process, 
(3) noisy skip-connection and temperature anneal (Noisy+Anneal): using both training techniques in the search process, 
(4) without noisy skip-connection and temperature anneal (Origin): searching the architecture without both training techniques. 
We use AUC and Logloss as metrics to evaluate the effectiveness of two types of training techniques.
Besides, we set the \textit{Origin} as the benchmark for calculating $\Delta_{AUC}$ and $\Delta_{Logloss}$.
As illustrated in Table \ref{tab:train-res}, \textit{Noisy} reduces the averaged Logloss $\Delta_{Logloss}$ by a large margin (0.0036), 
while \textit{Anneal} increases the averaged AUC $\Delta_{AUC}$ by 0.0017. 
This demonstrates that both injecting noise into skip-connection and annealing the softmax temperature can reduce the discrete loss over datasets with different scales.
The former can reduce the averaged Logloss significantly, while the latter can increase the averaged AUC remarkably. Note that the combination of these two techniques (\textit{Noisy+Anneal}) does not show superior performance than individual ones. We will conduct further investigation on this phenomenon in the future.
\subsection{Case Study (RQ4)}
\label{sec:ssa}
We run experiments of Avazu and Criteo multiple times with different random seeds. Since the obtained architectures identified by AutoPI are similar in each task, two representative architectures for Avazu and Criteo respectively are shown in Figure \ref{fig:searched} for more detailed analysis of our search space. We analyse the architecture searched via AutoPI from three perspectives: (1) The order of interactions. The interaction cell in Avazu consists of a self-attention (Att) and seven skip-connections, building a 
second-order feature interaction, of which order is lower than Crtieo. This demonstrates the Avazu relies more on lower-order cross features which 
is consistent with the observation in  \cite{cheng2020adaptive}.
(2) The type of operators. In our operation space, there exist various operators with different complexity. We observe that compared to the architecture searched for Avazu, more complex operators are selected by AutoPI for Criteo, which means our method can automatically generate larger networks for more complicated datasets.
(3) The integrated towers. An interesting observation is that the ensemble cell of Criteo has two repeated connections (i.e., $6, 7\rightarrow 8$ and $6, 7\rightarrow 10$), which can be simplified to a three-tower ensemble (i.e., integrating intermediate nodes $7, 8, 9$). This illustrates that Criteo only needs a small number of towers to be integrated because of its higher-order interactions. On the contrary, Avazu needs more towers to be integrated because it prefers lower-order interactions.

\section{Discussion}
\label{sec:disc}
In this paper, we propose a general method AutoPI to automatically search for both effective and efficient architectures for CTR prediction in different scenarios with different scales. 
Extensive experiments on a diverse suite of benchmark datasets demonstrate the model searched by our method outperforms both the hand-crafted models and other NAS methods in terms of AUC and Logloss. Coupling with the gradient-based search strategy, AutoPI is more computationally efficient than all available NAS methods.

To reduce the performance gap between the continuous architecture encoding and discrete architecture, both temperature anneal and noisy skip-connection are verified to be effective. Finally, in the case study, some interesting observations regarding to the architecture properties for large-scale and small-scale datasets are summarized, which can provide insights for researchers on how to design better CTR models.

\begin{acks}
This work was supported by NSFC under Grant No. 61936011 and No. 615210002, Beijing Key Lab of Networked Multimedia. Research reported in this publication was also supported by the National Library of Medicine of the National Institutes of Health, USA, under award number R01LM013151. We thank Tianchi Huang and Xin Yao for fruitful discussions. Finally, life is fine.
\end{acks}

\bibliographystyle{ACM-Reference-Format}
\balance
\bibliography{sample-base}










\end{document}